\documentclass[12pt]{article}
\usepackage{amssymb}
\usepackage{bbold}
\usepackage{graphicx}

\begin{document}
\title{Supernova Brightening from Chameleon-Photon Mixing}
\author {C. Burrage\footnote{C.Burrage@damtp.cam.ac.uk}\\
Department of Applied Mathematics and Theoretical Physics,\\
Centre for Mathematical Sciences,\\
Cambridge,\\
CB2 0WA, UK}

\maketitle
\abstract{Measurements of standard candles and measurements of standard rulers  give an inconsistent picture of the history of
  the universe.  This discrepancy can be explained if photon number is not
conserved as computations of the luminosity distance must be
modified.  I show that photon number is not conserved when photons
mix with chameleons in the presence of a magnetic field.
The strong magnetic fields in a supernova mean that the probability of
a photon converting into a chameleon in the interior of the supernova
is high, this results in a large flux of chameleons at  the surface of the
supernova.  Chameleons and photons also mix as a result of the intergalactic magnetic
field.   These two effects combined cause the image of the supernova
to be brightened resulting in a model which fits both observations of
standard candles and observations of standard rulers.}

\bibliographystyle{unsrt}
\section{Introduction}
Observations of standard candles  and of standard rulers allow us to
learn how the universe has evolved in recent times.  Both sets of observations favour
a universe containing a large component of dark energy but the
cosmologies preferred by the two sets of data are not consistent; in
fact at higher redshifts they appear to diverge.  This indicates that
there is still something missing in our understanding of the
accelerated expansion of the universe.  In this paper I will show how
the discrepancy between the two sets of observations can be explained
if photon number is not conserved because photons are allowed to mix
with chameleon particles.

In metric theories of gravity where photons
travel on unique null geodesics there is a well-defined notion of distance as
long as photon number is conserved.
The luminosity distance, $d_L(z)$, to an object at redshift $z$  and the angular diameter
distance, $d_A(z)$, to the same object are related by the reciprocity
relation \cite{Etherington}
\begin{equation}
\label{duality}
d_L(z)=d_A(z)(1+z)^2
\end{equation}
If photon number is not conserved then this relation no longer
holds and a discrepancy in the
distance measures should be observed.  There are a number of current
proposals which do not conserve photon number; the light from distant objects could be scattered
by dust or free electrons in the intergalactic medium, photons may
decay,  or the photon
could mix with another light state such as the axion.

It had been thought that loss of photons in the intergalactic medium could explain the observed
dimming of type 1a supernova without the need for dark
energy  \cite{Csaki:2001yk,Song:2005af,Kaloper:2007gq}. Photon number non-conservation changes the luminosity distance
to an object but does not affect distance measurements from standard
rulers.  Therefore, if loss of photons is the explanation for the
dimming of supernova, angular diameter distance measurements should be consistent with an $\Omega_{\Lambda}=0$ universe.
This is not the case; observations of standard rulers imply a universe with a significant dark energy
component.

\begin{figure}
\begin{center}
\includegraphics[width=14cm,height=8cm]{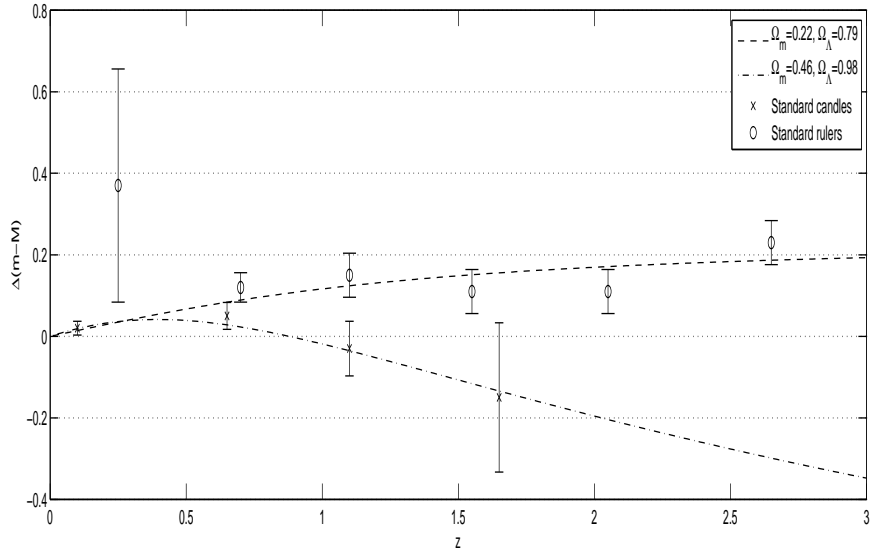}
\caption{Distance modulus [$m-M=5\log(d_L)-25=5\log(d_A(1+z)^2)-25$]
  compared to a $\Lambda$CDM universe plotted against redshift,
  showing the best fits to observations of standard candles and
  standard rulers.}
\label{fig:DeltaM}
\end{center}
\end{figure}

When luminosity distances and angular diameter distances  are compared there is a noticeable
disagreement at $z \gtrsim0.5$ \cite{Bassett:2003vu,Bassett:2004}.  This is shown in Figure
\ref{fig:DeltaM} where distance modulus (compared to a $\Lambda$CDM
universe) is plotted against redshift for measurements
coming from standard rulers and standard candles.  The standard
candles used are type 1a supernovae
\cite{Riess:2004nr,Tonry:2003zg,Barris:2003dq,Knop:2003iy} and the
standard rulers used are FRIIb radio galaxies \cite{Daly:2003iy,Daly:2002kn}, compact radio
sources \cite{Gurvitis,Jackson:2006ib,Jackson:2003jw} and X-ray
clusters \cite{Allen:2002eu}. Considering the two sets of data
separately the best fit
to the angular diameter distance data is a universe with $\Omega_m=0.22$,
$\Omega_{\Lambda}=0.79$ but the best fit to the luminosity distance data is a very closed universe
with $\Omega_m=0.46$, $\Omega_{\Lambda}=0.98$, such a universe is already ruled out by
measurements of the CMB.  If the explanation for this apparent
violation of the reciprocity relation is that photon number is not
conserved, then computed luminosity distances need to be modified.

Photon number non-conservation does not affect the angular diameter
distance to an object so the cosmology which best fits observations of
standard rulers is the correct one.  If this is the case Figure \ref{fig:DeltaM}
implies that non-conservation of photon number actually brightens the
image of a supernova.  This is in contrast to all previously studied
mechanisms for non-conservation of photon number which all predict
that photons should be lost in the intergalactic medium, and thus that
the supernova is dimmed \cite{Bassett:2003vu,Bassett:2004}.
In this paper I show that the images of supernova can be brightened if
photons are allowed to mix with chameleons.

The chameleon is a scalar particle which arises in certain models of
scalar-tensor gravity \cite{Khoury:2003rn,Khoury:2003aq}.  In the
Einstein frame the action is
\begin{eqnarray}
S&=&\int d^4x\sqrt{-g}\left(\frac{M_P^2R}{2}-\frac{1}{2}(\partial
\phi)^2-V(\phi)\right)\nonumber\\
& &-\int d^4x \mathcal{L}_m(\psi_m^{(i)},g_{\mu,
  \nu}^{(i)})
\end{eqnarray}
where $\psi_m^{(i)}$ are matter fields which couple to the metric $g^{\mu
  \nu}_{(i)}=e^{-2\phi/M_i}g^{\mu \nu}$.  Assuming that the
  chameleon couples to all forms of matter in the same way so that
  $M_i=M$ $\forall i$ the current bound on $M$ is $10^6\mbox{
  GeV}<M$ \cite{Mota:2006fz}. I assume a runaway potential of the form
\begin{equation}
\label{pot}  
V(\phi)=\Lambda^4e^{\Lambda^n/\phi^n}
\end{equation}
but the chameleon feels an effective potential
\begin{equation}
V_{eff}(\phi)=V(\phi)+\rho e^{\phi/M}
\end{equation}
which depends on the local matter density $\rho$.  The chameleon sits
 in the minimum of the effective potential with a mass which depends
 on the local matter density.  In high
  density regions such as on earth the chameleon becomes very massive
 and so evades current experimental bounds, but in low density regions
 it can become almost massless.  The way in which the scalar field can
 change its mass is known as  the `chameleon mechanism' and this 
  allows the scalar field to evade all current searches for
 fifth-force effects or violations of the equivalence principle if  $\Lambda \sim 10^{-3}\mbox{ eV}$. In addition it has been shown that the chameleon is compatible with recent searches for
  axion like particles at PVLAS and CAST
 \cite{Brax:2007hi,Brax:2007ak}.  

As the universe evolves the local energy density falls and the
chameleon changes its properties as the universe cools.  In particular
at early times the chameleon behaves like
dust but as the universe cools below  temperatures $T \approx 10\mbox{
  MeV}$ the chameleon  begins to behave like a fluid with equation of state
$\omega=-1$ \cite{Brax:2004qh,Brax:2004px}.
In this way the chameleon is a natural
  explanation for dark energy.

The paper is organised as follows.  Section \ref{homogeneous} shows
how photons mix with
chameleon particles and gives the probability of a photon converting into a
chameleon in a homogeneous magnetic field.  Section \ref{varying} discusses the
probability of conversion as the particles pass through a magnetic
field made up of many randomly oriented domains and section
\ref{supernova} explains how photons can be converted into chameleons
in the interior of a supernova, resulting in a flux of chameleons at
the surface.  Then Section \ref{sec:brightening} describes how
this can account for the brightening of the image of the
supernova. In Section \ref{CMB} I show that chameleon-photon mixing is in agreement with observations of the CMB.   I
conclude in section \ref{conc}.

\section{Photon-Chameleon Mixing}
\label{homogeneous}
 In
the presence of a magnetic field the chameleon couples to
photons. Assuming that the coupling of the chameleon to matter is
universal, the interaction term in the Lagrangian is
\begin{equation}
\mathcal{L}_{int} =\frac{\phi B^2}{M}
\end{equation}
 In a homogeneous
magnetic field where the particles propagate along the $x$ axis which is aligned in the direction of the magnetic field the equations governing
the evolution of the fields \cite{Raffelt:1987im} are 
\begin{equation}
\left[\omega^2 +\partial^2_x+\left(\begin{array}{ccc}
-\omega_p^2 & 0 & 0\\
0 & -\omega_p^2 & \frac{B\omega}{M}\\
0 & \frac{B\omega}{M} & -m_c^2
\end{array}\right)\right]\left(\begin{array}{c}
A_{\parallel}\\
A_{\perp}\\
\phi
\end{array}\right)=0
\end{equation}
where $\omega$ is the frequency of the photons and  $B$ is
the magnitude of the magnetic field. As photons and chameleons propagate through the
intergalactic medium they pass through a plasma of ionised electrons
\cite{Csaki:2001jk}. $\omega_p$ is the
frequency of this plasma; $\omega_p^2=4\pi \alpha n_e/m_e\approx
10^{-47}\mbox{ GeV}^2$, where  $n_e$ is
the electron number density and $m_e$ is the electron mass. 
$A_{\parallel}$ and $A_{\perp}$ are the polarisations of the photon
 parallel and perpendicular to the magnetic field and $\phi$ is the chameleon
field.  Note that the
chameleon only  mixes with the polarisation of the photon which is
orthogonal to the magnetic field. $m_c$ is the chameleon mass which, given the potential (\ref{pot}), is
\begin{equation}
m_c^2=n(n+1)\frac{\Lambda^{4+n}}{\phi_{min}^{2+n}}
\end{equation}
where 
\begin{equation}
\phi_{min}=\left(\frac{n\Lambda^{4+n}M}{\rho}\right)^{\frac{1}{1+n}}
\end{equation}
is the value of $\phi$ at the minimum of the effective potential.  In
what follows I will assume $n=\mathcal{O}(1)$.

The probability of an orthogonally polarised photon converting into a chameleon whilst
travelling a distance $x$ through this homogeneous field is
\begin{eqnarray}
P(x)&=&\frac{4\omega^2B^2}{M^2(\omega_p^2-m_c^2)^2+4\omega^2B^2}\nonumber\\
& &\times\sin^2\left(\frac{x\sqrt{M^2(\omega_p^2-m_c^2)^2+4\omega^2B^2}}{4\omega
  M}\right)
\label{PIGM}
\end{eqnarray}
The observed  dimming of type 1a
supernova is achromatic.  hence if chameleon-photon mixing modifies
observations of supernovae the effect must also be achromatic.  If the
particles pass through a magnetic domain of size $L_{dom}$ there are two regimes in which the conversion
probability (\ref{PIGM}) is independent of frequency:  In the limit of high energy photons $M|\omega_p^2-m^2| \ll B\omega$ the
mixing is maximal and independent of the photon energy.
\begin{equation}
\label{probIGM}
P\approx \sin^2\left(\frac{L_{dom}B}{2M}\right)
\end{equation}
Alternatively if the oscillation length
\begin{equation}
L_{osc}=\frac{4\pi \omega
  M}{\sqrt{M^2(\omega_p^2-m_c^2)^2+4\omega^2B^2}}
\end{equation}
is much greater that the size of a magnetic domain $2\pi L_{dom} \ll
L_{osc}$ then the probability of conversion (\ref{PIGM}) is
\begin{equation}
\label{probS}
P \approx \frac{B^2L_{dom}^2}{4M^2}
\end{equation}
which is independent of photon energy.  It should be noted that both
(\ref{probIGM}) and (\ref{probS}) are also independent of the mass of the
chameleon.  

Current observations suggest that the
  intergalactic magnetic field has coherence length $L_{dom} \sim
  1\mbox{ Mpc}$ and  magnitude $B\lesssim
  10^{-9}\mbox{ G}$
  \cite{Kronberg:1993vk,Blasi:1999hu,Jedamzik:1999bm}.  The
  density of the intergalactic medium is $\rho_{IGM} \sim
  10^{-44}\mbox{ GeV}^4$ so the mass of the chameleon is
  $m_c^2 \lesssim 10^{-45}\mbox{ GeV}^2$. An optical photon $\omega \approx
  10\mbox{ eV}$ passing through the
  intergalactic medium  is in the high energy regime if
\begin{equation}
M\lesssim 10^{10}\mbox{ GeV}
\end{equation}
so that the
  probability of conversion is given by (\ref{probIGM}).
For this probability to be small requires  $BL_{dom}<2M$.

\subsection{Conversion in a Varying Background}
\label{varying}
The intergalactic magnetic field is not homogeneous on large scales; it is  made up of many randomly orientated
magnetic domains. To calculate the probability of a photon converting into
a chameleon whilst travelling through the intergalactic medium I
assume that:  Particles
traverse $N$ domains of equal length,  $B$ is homogeneous in each
domain and there is a discrete change in $B$ from one domain to
another.  The component of the magnetic field parallel to the
direction of flight has a random orientation but equal size in each
domain.

The initial state is
\begin{equation}
\alpha_1(0)|\gamma_1\rangle +\alpha_2(0)|\gamma_2\rangle +\alpha_c(0)|c\rangle
\end{equation}
where $|\gamma_i\rangle$ are the photon states parallel and perpendicular to
the magnetic field in the first domain  and $|c\rangle$ is the
chameleon.  The initial photon and
chameleon fluxes are
\begin{eqnarray}
I_{\gamma}(0) &\sim& |\alpha_1(0)|^2+|\alpha_2(0)|^2\\
I_{c}(0) &\sim& |\alpha_c(0)|^2
\end{eqnarray}
In the n-th domain the magnetic field is tilted by an angle $\theta_n$
compared to the first domain, so that
\begin{eqnarray}
|\gamma^n_{\parallel}\rangle&=&\cos\theta_n|\gamma_1\rangle+\sin\theta_n|\gamma_2\rangle\\
|\gamma^n_{\perp}\rangle&=& -\sin\theta_n|\gamma_1\rangle+\cos\theta_n|\gamma_2\rangle
\end{eqnarray}
and the transition probability $P$ in each domain is given by (\ref{PIGM}). 
Assuming
that $P$ is small and that
$\theta_n$ is a random variable so that on average $\cos^2\theta_n \sim
\sin^2\theta_n \sim 1/2$ then at the
end of the n-th domain
\begin{eqnarray}
I_c(y) &=&
\frac{1}{3}(I_c(0)+I_{\gamma}(0))+\frac{Q(y)}{3}(2I_c(0)-I_{\gamma}(0))\label{Ic}\\
I_{\gamma}(y) &=&
\frac{2}{3}(I_c(0)+I_{\gamma}(0))+\frac{Q(y)}{3}(I_{\gamma}(0)-2I_c(0))\label{Igamma}
\end{eqnarray}
where
\begin{equation}
\label{Q}
Q(y)=\left( 1-\frac{3P}{2}\right)^{y/L_{dom}}
\end{equation}
and $y(z)$ is the
proper particle distance to the astronomical object.  If $P$ is small then at large distances $Q(y)$ becomes
exponentially small and the system reaches an equilibrium
configuration with a third of the initial flux in chameleons and two
thirds in photons. 

The probability of a single photon converting to a chameleon in a distance
$y$ is
\begin{eqnarray}
P_{\gamma \rightarrow c}(y)&=&\frac{1}{3}(1-Q(y))\\
&\lesssim & \frac{y^2B^2}{8M^2N}
\label{totalprob}
\end{eqnarray}
It can be seen from (\ref{Ic}), (\ref{Igamma}) that if the ratio of the initial chameleon flux to the initial photon
flux is large enough more photons will be received than were emitted; the image of the supernova is brightened.

\subsection{Conversion in the Supernova}
\label{supernova}
A flux of chameleons is emitted by a supernova if some photons are
converted into chameleons in the interior of the supernova.
A type 1a supernova is thought to be the thermonuclear explosion of
 a white dwarf whose mass is close to the Chandrasekhar limit.  To
 compute the probability of conversion between photons and chameleons
 inside the supernova I consider a simple model: The supernova is a sphere of uniform density with initial
 radius $R_0 \sim 10^9\mbox{ cm}$.  The supernova expands with outer velocity
 $v=c/30 \sim 10^9 \mbox{ cm/s}$.  I also
assume that the size of a magnetic domain in the
supernova is roughly
 equal to the length of the mean free path of the photons $L_{dom}
 \approx L_{mfp}$.  I assume that only photons are produced by the reactions driving the
explosion of the supernova and that the photons are emitted uniformly
 throughout the volume of the supernova.  Peak luminosity occurs about
10 days after the start of the explosion. 

 The explosion of a supernova is
 homologous and so the magnetic field obeys
\begin{equation}
\frac{B_{SN}(t)}{B_{WD}}=\left(\frac{R_{WD}}{R_{SN}(t)}\right)^2
\end{equation}
where $B_{SN}(t)$ and $R_{SN}(t)$ are the magnetic field and radius of the
supernova at time $t$ after the start of the explosion and  $B_{WD}$ and
$R_{WD}$ are the magnetic field and radius of the initial white dwarf.
Models of the magnetic field of a white dwarf vary, but all
predictions lie in the range $10^5\mbox{ G} \lesssim B_{WD} \lesssim
10^{11}\mbox{ G}$.  The size of
the mean free path of photons at peak luminosity of the supernova is
also not well known but is expected to satisfy $10^6\mbox{ cm}
\lesssim L_{mfp} \lesssim 10^{14}\mbox{ cm}$.  
At peak luminosity the mean free path of photons is much smaller than
the radius of the supernova and  the path of the photons can be modelled
as a random walk so that it takes the photon
$N=3R^2/L_{mfp}^2$ steps to escape from a region of radius $R$.   

Inside the supernova the chameleon is more massive than in the
intergalactic medium, however because
the oscillation length of optical
photons is much greater than the coherence length of the magnetic
field the oscillations are still independent of frequency (\ref{probS}). The probability of
a photon converting into a chameleon inside the supernova is thus
\begin{eqnarray}
P_{\gamma \rightarrow c}(R_{SN}) &\lesssim& \frac{3B_{SN}^2R_{SN}^2}{8M^2}\\
& \lesssim& 9.4\times 10^{32}
  \left(\frac{B_{WD}^2}{\mbox{GeV}^2M^2}\right)
\end{eqnarray}
Therefore even though only photons are produced by the
thermonuclear reactions in the supernova, the relatively high
probability for a photon to
convert into a chameleon in the interior of the supernova means
that there is a significant flux of chameleons at the surface of the supernova.

In \cite{Grossman:2002by} the possibility of producing a flux of axions from the
supernova was considered in a similar way to that described above.  In the photon-axion coupling
model $M\approx 10^{11}\mbox{ GeV}$ which means that the probability
of conversion in the supernova is negligible.  The chameleon
mechanism, which changes the mass of the scalar field, means that the
experimental constraints on $M$ in the chameleon model are less
severe.  $M$ can be much smaller than in the axion model and therefore
the probability of conversion can be much higher.

\section{Supernova Brightening}
\label{sec:brightening}
If there is a flux of chameleons at the surface of the supernova, the
ratio of the flux of photons received on earth to the flux of photons
leaving the supernova is 
\begin{eqnarray}
 P_{\gamma \rightarrow
 \gamma}(y)&=&\frac{I_{\gamma}(y)}{I_{\gamma}(0)}\nonumber\\
&=&\frac{2}{2+(1-\frac{3}{2}P_{SN})^N}+Q(y)\left(\frac{(1-\frac{3}{2}P_{SN})^N}{2+(1-\frac{3}{2}P_{SN})^N}\right)\label{brightening}
\end{eqnarray}
where $y=0$ is now the surface of the supernova.  $P_{SN}$ is the
probability of conversion in one domain in the supernova and $Q(y)$ is
given in (\ref{Q}) with $P=P_{IGM}$ the probability of conversion in
one domain in the intergalactic magnetic field.
If photon number is not conserved in the intergalactic medium then the reciprocity relation between
luminosity distance and angular diameter distance
(\ref{duality}) must be modified by sending $d_L
\rightarrow d_L/\sqrt{P_{\gamma \rightarrow \gamma}}$. 

\begin{figure}
\begin{center}
\includegraphics[width=14cm,height=8cm]{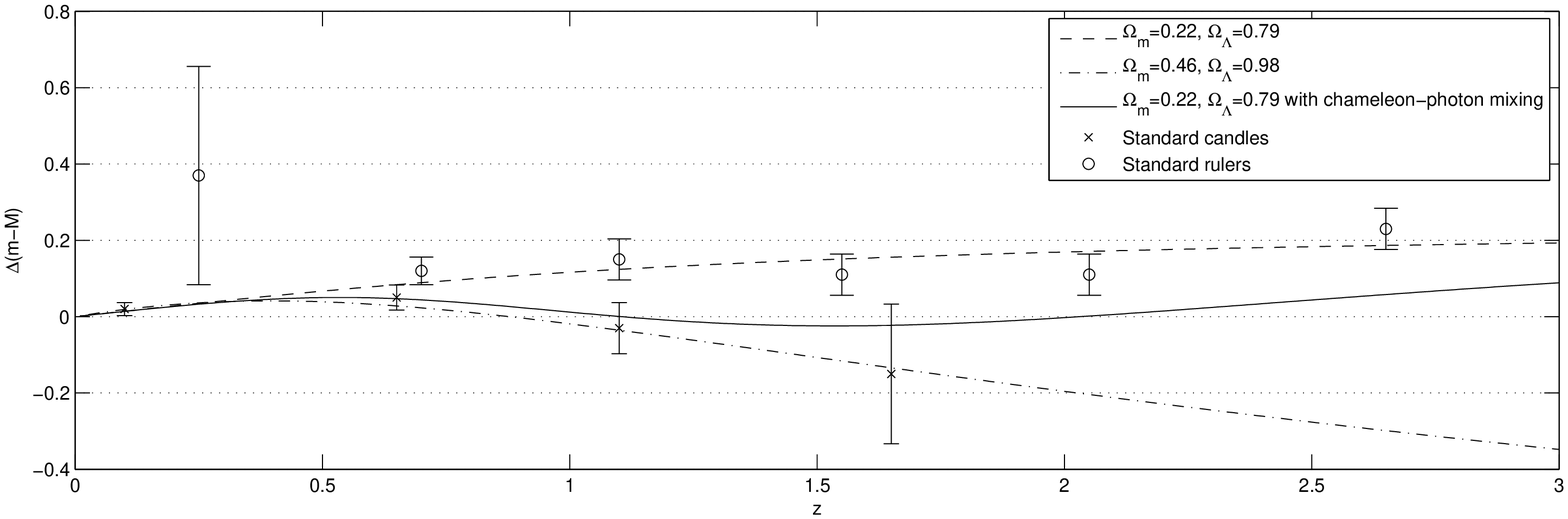}
\caption{  Assuming standard rulers give the correct relation between
  $\Delta(m-M)$ and $z$, the solid line shows the prediction for
  observations of type 1a supernova if the probability of photon
  survival has the form (\ref{form}) with
  $A= 0.99993$ and $c= -0.056$.}
\label{fig:brightening2}
\end{center}
\end{figure}

\begin{figure}
\begin{center}
\includegraphics[width=14cm,height=8cm]{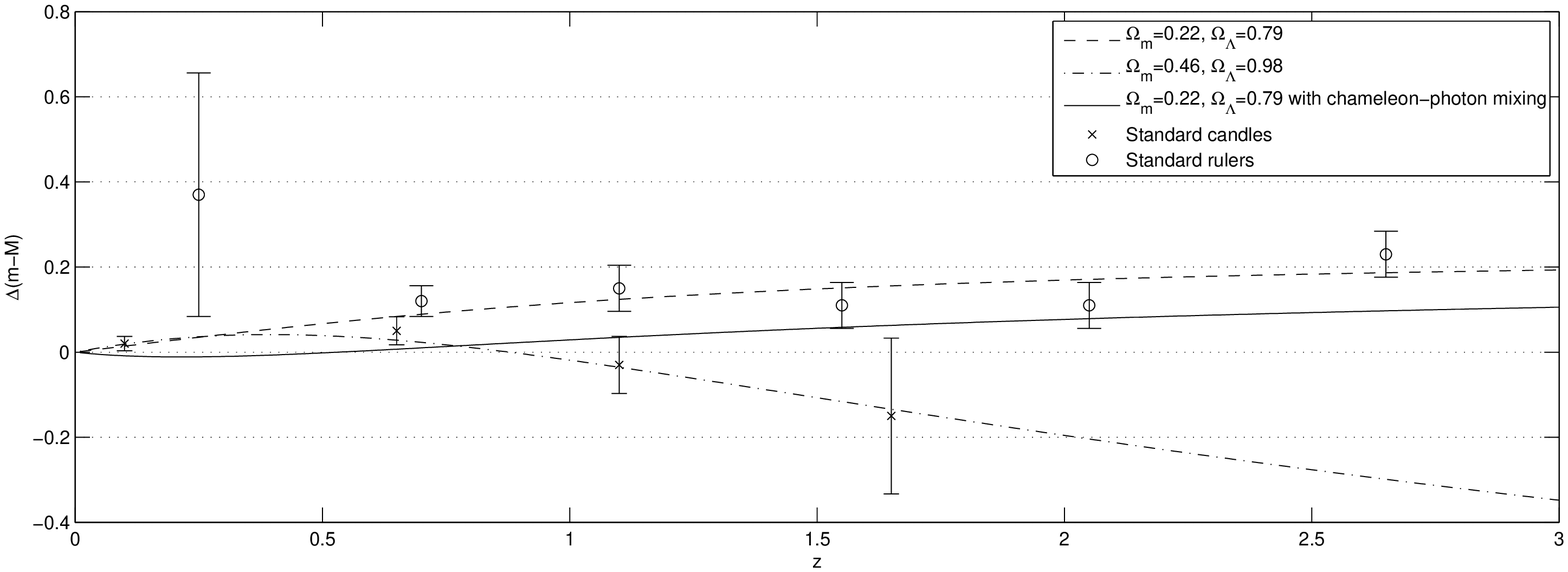}
\caption{Assuming standard rulers give the correct relation between
  $\Delta(m-M)$ and $z$, the solid line shows the prediction for
  observations of type 1a supernova if the probability of photon
  survival has the form (\ref{form}) with
  $A= 1.271$ and $c= 1.10$.}
\label{fig:brightening}
\end{center}
\end{figure}

Writing the photon survival probability as
\begin{equation}
\label{form}
P(z)=A+(1-A)e^{-y(z)H_0/c}
\end{equation}
where $A$ and $c$ are real constants.  I consider the effect of
chameleon-photon mixing on predictions for supernova observations.  As
chameleon-photon mixing has no effect on the angular diameter
distance I  assume that measurements of standard rulers  give the correct relation between distance and
redshift.   It is possible to fit the
observed supernova well if $c<0$.  The example in Figure
\ref{fig:brightening2} has $1-A=7\times 10^{-5}$ and $c=-0.056$.  However
$c<0$ corresponds to $\ln|1-3P_{IGM}/2|>0$ which is not possible
within this model.  If $c>0$ the tension between the $d_A$ and $d_L$
measurements can be eased by photon-chameleon mixing.  Suitable values
of the parameters are $A=1.271$ and $c=1.10$ and the resulting
prediction for the supernova is shown in Figure \ref{fig:brightening}. For $z\lesssim 0.5$
the prediction is close to a $\Lambda$CDM universe and for larger $z$
there is a constant brightening from the radio galaxy data.  Clearly
more observations of high redshift supernova
or smaller error bars would significantly improve the constraints on
our model.

The values of $A$ and $c$  used in Figure \ref{fig:brightening} correspond to $P_{SN}N \approx 0.95$
and $P_{IGM} \approx  10^{-4}$ where
\begin{equation}
P_{SN}N \approx \frac{3B_{WD}^2R_{WD}^4}{4M^2R^2_{SN}}, \;\;\;\;\; 
P_{IGM}  \approx \frac{B_{IGM}^2L_{dom}^2}{4M^2}
\end{equation}
This implies the following relations between
parameters
\begin{equation}
B_{WD}\sim 10^3\left(\frac{M}{\mbox{GeV}}\right)\mbox{ G},\;\;\;\;\;
B_{IGM}\sim 10^{-20}\left(\frac{M}{\mbox{GeV}}\right)\mbox{ G}
\end{equation}
which are consistent with all current experimental bounds.

In these calculations I have neglected the effect of the galactic magnetic field.  The
probability of conversion in one domain  (\ref{probIGM}) depends on the combination
$BL_{dom}$.  The galactic magnetic field has strength $B_G \approx
10^{-5}\mbox{ G}$ and coherence length $L_{dom}=100\mbox{ pc}$ so
that $B_GL_{dom}\approx 10^9\mbox{ GeV}$.  This is
the same order of magnitude for $BL_{dom}$ as for an intergalactic magnetic field
with $B_{IGM}\approx10^{-9}\mbox{ G}$ and coherence length
$L_{dom}\approx 1\mbox{ Mpc}$.  So in this case passing from the
intergalactic magnetic field to the galactic magnetic field does not
affect the probability of conversion.  If $B_{IGM} \ll 10^{-9} \mbox{
    G}$  then there will be a change in the
probability of conversion when the photons move into the galaxy,
however the distance travelled through the galaxy is small compared to
the total distance from the supernova so the effects of the galaxy
will at most be a small correction on the above result.

\section{CMB Photons}
\label{CMB}
 If
microwave photons from the CMB oscillate into chameleon states in the  intergalactic magnetic
field their anisotropy could be large due to variations in the
magnetic field and thus disagree with observations.  The frequency of a photon from the CMB is $\omega \sim
10^{-4}\mbox{ eV}$ so microwave photons fall into the low energy
regime  $M|\omega_p^2-m^2| \gg B\omega$  where the mixing is small
and the probability of chameleon-photon mixing is bounded by 
\begin{eqnarray}
P &\leq&
4B^2\omega^2/M^2(\omega_p^2-m^2)^2\\
&\lesssim& 10^{-6}
\end{eqnarray}
 for all allowed
values of $M$. The anisotropy sourced by one magnetic domain is less
that the primordial CMB anisotropy $\Delta T/T \sim 10^{-5}$.

If intergalactic magnetic fields are a relatively
recent phenomenon and only exist out to redshifts of a few then  the photons from the
CMB have not travelled far enough through a magnetic field to reach their
equilibrium configuration. The probability of conversion is not significantly
enhanced by travelling through the many magnetic domains in the
intergalactic medium and therefore chameleon-photon mixing does not conflict with
observations of the CMB. The effect on the CMB of axion-photon mixing
was considered in \cite{Mirizzi:2005ng} but because we do not require a
large probability of conversion in the intergalactic medium we are
able to avoid the strict bounds that observations of the black body
spectrum of the CMB put on the axion photon mixing model.

  An observation of a primordial magnetic field would put
severe constraints on the chameleon model because then photons from
the CMB would travel for large distances through a magnetic field.
Conversely if the existence of chameleons is demonstrated, observations of the CMB would put strict bounds on the existence of a primordial magnetic field.

\section{Conclusions}
\label{conc}
The chameleon model provides an explanation for the observed
accelerated expansion of the universe.  Observations of
standard candles and standard rulers do not give a consistent picture
of this acceleration which implies that the reciprocity relation
(\ref{duality}) does not hold.  This can be explained if the chameleon couples to photons in the presence of
a magnetic field.  The strong magnetic field inside a supernova means
that chameleons and photons mix in the interior of the
supernova and so there is a flux of chameleon particles at the surface of the
supernova.  This is in contrast to axion-photon coupling which does
not allow for a flux of axions at the surface of the supernova. Photons and chameleons also mix as they travel to earth
through the intergalactic magnetic field.  
This requires $M\lesssim10^{10}\mbox{ GeV}$ to ensure the effect of
the mixing 
is achromatic for optical photons.  A chameleon model with a coupling
of this strength should be detectable in future experiments looking for
chameleonic afterglow \cite{Gies:2007su,Ahlers:2007st} or Casimir
forces.  

The overall effect of mixing between photons and chameleons 
is that observers on earth see a brightened image of the supernova.
The brightening of supernovae eases the tension between
observations of standard candles and standard rulers.  Future observations of high redshift
supernova will significantly improve the constraints on this model.
\section*{Acknowledgements}
I would like to thank A.C. Davis, P. Brax and D. Shaw for very useful
discussions.  This work was supported by STFC.

\bibliography{chameleon}
\end{document}